\documentclass[aps,prl,showpacs,twocolumn,floats,superscriptaddress]{revtex4}
\usepackage{amsmath}
\usepackage{bm}
\usepackage{graphicx}
\usepackage{epstopdf}
\DeclareMathOperator{\Tr}{Tr}
\newcommand{\parent}[1]{\left( #1\right)}
\newcommand{\abs}[1]{\left| #1\right|}
\newcommand{\bra}[1]{\left< #1\right|}
\newcommand{\ket}[1]{\left| #1\right>}
\newcommand{\bracket}[2]{\left< #1\: \right| \left. #2\right>}

\usepackage{color}
\newcommand{\nc}{\newcommand}
\nc{\rr}{\color{red}}
\nc{\bb}{\color{blue}}
\nc{\gr}{\color{green}}

\begin{document}

\title{Displacement Echoes: Classical Decay and Quantum Freeze}
\author{Cyril Petitjean}
\affiliation{D\'epartement de Physique Th\'eorique,
Universit\'e de Gen\`eve, CH-1211 Gen\`eve 4, Switzerland}
\author{Diego V. Bevilaqua}
\affiliation{Department of Physics, Harvard University, Cambridge, MA 02138}
\author{Eric J. Heller}
\affiliation{Department of Physics, Harvard University, Cambridge, MA 02138}
\affiliation{Department of Chemistry and Chemical Biology, Harvard University, Cambridge, MA 02138}
\author{Philippe Jacquod}
\affiliation{Physics Department, 
   University of Arizona, Tucson, AZ 85721}

\date{\today}

\begin{abstract}
Motivated by neutron scattering experiments, 
we investigate the decay of the fidelity with which a wave packet is 
reconstructed 
by a perfect time-reversal operation performed after a phase space 
displacement. In the semiclassical limit, we show that 
the decay rate is generically given by the Lyapunov exponent of the classical 
dynamics. For small displacements, we additionally show that,  
following a short-time Lyapunov decay, the decay freezes well above the 
ergodic value because of quantum effects.
Our analytical results are corroborated by numerical simulations.
\end{abstract}


\maketitle

The fidelity with which a wavefunction is reconstructed
after an imperfect time-reversal operation 
was originally introduced as a measure of reversibility in
quantum mechanics~\cite{peres}. 
Dubbed the Loschmidt Echo, it
has received much attention in recent years in the context
of decoherence and the quantum classical 
correspondence~\cite{jalabert,jacquod,tomsovic,karkuszewski,vanicek,gorin}. For a generic 
perturbation of the Hamiltonian, four different decay regimes were
found: the Gaussian perturbative regime, 
the Fermi Golden Rule (FGR) regime, the Lyapunov regime, 
and the regime of classically large perturbations.
Of special interest is the Lyapunov regime where the purely quantum mechanical
fidelity decays with the Lyapunov exponent of the classical dynamics. It
suggests the existence of a universal regime of environment-independent 
decoherence rate~\cite{jalabert,zurek,cyril}.

In this letter we analyze the decay of the Loschmidt echo under a new,
non-generic perturbation, namely an instantaneous phase space displacement. 
Our investigation is partly inspired  by spectroscopies such as neutron scattering, M\"ossbauer $\gamma$-ray,  and certain electronic transitions in molecules and solids~\cite{displacement-spectroscopy,caveat1,ins}.
In these  spectroscopies phase space displacement (momentum boost or position shift) takes place with little or no change in the potential. Under this non-generic 
perturbation, we find that the decay rate of the average fidelity 
is always set by the Lyapunov exponent. Moreover, for small displacements,
the initial Lyapunov decay is followed at larger times by a quantum freeze of 
the fidelity at a displacement-dependent saturation value. Both 
semiclassics and random matrix theory predict that the freeze persists 
up to infinitely large times.

As our starting point, we recall that 
the differential cross section for incoherent 
neutron scattering and M\"ossbauer emission/absorption can be calculated from the following correlation function \cite{lovesey,vanhove} (from now on we set $\hbar \equiv 1$)
\begin{eqnarray}
Y_{jj}\parent{\bf{P},t}=\left<  e^{-i\bf{P}\cdot \bf{\hat{r}}_j}e^{i\hat{H}t}e^{i\bf{P}\cdot \bf{\hat{r}}_j}e^{-i\hat{H}t}\right>.
\end{eqnarray}
Here, the brackets represent an ensemble average, $\bf{\hat{r}}_j$ are the position 
operators of the nuclei and $\hat{H}$ is the typical Hamiltonian of the 
target system. 
The ensemble average of the correlation function can be written~\cite{ins}:
\begin{eqnarray}
Y_{jj}\parent{\bf{P},t}& \approx &\frac{1}{Q}\int \parent{\frac{d^{2N}\alpha}{\pi^N}}\Phi\parent{\alpha} \nonumber \\
& &\times \bra{\alpha}e^{-i\bf{P}\cdot \hat{\bf{r}}_j}e^{i\hat{H}t}e^{i\bf{P}\cdot \hat{\bf{r}}_j}e^{-i\hat{H}t}\ket{\alpha}, \label{auto}
\end{eqnarray}
where $\ket{\alpha}$ are coherent states with $N$ degrees of freedom, $Q=\Tr \left[ e^{-\beta \hat{H}}\right]$, and $\Phi\parent{\alpha}$ is a thermal weight, which tends to  $e^{-\beta H_{cl}\parent{\alpha}}$ at high temperatures. 
The notation $e^{i\hat{H}_{\bf{P}}t}=e^{-i\bf{P}\cdot \hat{\bf{r}}_j}e^{i\hat{H} t}e^{i\bf{P}\cdot \hat{\bf{r}}_j}$ suggests that we
identify the kernel of the integral $I\parent{t}=\bra{\alpha}e^{-i\bf{P}\cdot \hat{\bf{r}}_j}e^{i\hat{H}t}e^{i\bf{P}\cdot \hat{\bf{r}}_j}e^{-i\hat{H} t}\ket{\alpha}$ with the kernel of a 
Loschmidt echo problem, we thus introduce
the momentum {\it displacement echo}
\begin{equation}\label{decho}
M_{\rm D}\parent{t}=\vert I(t)\vert^2 = \big|\bra{\alpha}e^{i\hat{H}_{\bf{P}}t} e^{-i\hat{H}t}\ket{\alpha}\big|^2.
\end{equation}

As introduction to our semiclassical calculation of the 
displacement echo $M_{\rm D}(t)$,
we first discuss the validity of the
diagonal approximation used in Ref.~\cite{jalabert} for the semiclassical approach
to the Loschmidt echo and
point out why this approximation is even better for the displacement echo.
The diagonal approximation for the Loschmidt echo equates each classical trajectory
$\gamma$ generated by an unperturbed Hamiltonian $H$ with a classical trajectory
$\gamma_V$ generated by a perturbed Hamiltonian $H_V=H+V$.
This procedure is not justified a priori in chaotic systems where one expects that
an infinitesimally small perturbation generates trajectories
diverging exponentially fast away from their unperturbed counterpart. 
It was however pointed out by Cerruti and Tomsovic~\cite{tomsovic}, and later by Vani\v{c}ek and 
Heller~\cite{vanicek}, that structural stability theorems~\cite{sst,vanicek2} justify this 
approximation. Roughly speaking one can show that,
given a uniformly hyperbolic Hamiltonian system $H$, and a generic 
perturbation $V$,
each classical trajectory $\gamma_V^{\prime}$ generated by the (still 
hyperbolic) 
perturbed Hamiltonian $H+V$ is almost always arbitrarily close to one 
unperturbed
trajectory $\gamma$. 
In general the two trajectories do not share common endpoints,
however these endpoints are close enough that they
are not resolved quantum-mechanically.
This is illustrated in the left panel of Fig.~1.
The semiclassical expression for the kernel of the Loschmidt echo involves a 
double sum
over both the perturbed and the unperturbed classical trajectories,
so that both $\gamma_V^{\prime}$ and $\gamma$ are included. After a stationary 
phase condition, this double sum is reduced to a single sum where
$\gamma_V^{\prime}$ and $\gamma$ are equated. In other words, a semiclassical 
particle follows $\gamma$ in the forward direction, and 
$\gamma_V^{\prime}$ in the backward direction because this is 
the best way to minimize 
the action. The action difference is simply given by the integral of the 
perturbation along the backward
trajectory, and it is in general time-dependent.

\begin{figure}[h]
\vspace{-3mm}
\begin{center}
\resizebox{!}{4.5cm}{\includegraphics{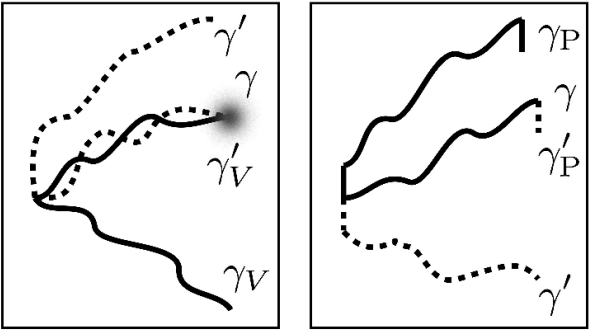}}
\end{center}
\vspace{-3mm}
\caption{Illustrative view of structural stability. Left panel: 
generic perturbation. $\gamma$ and $\gamma'$ are two orbits of the 
unperturbed Hamiltonian, $\gamma_V$ is the orbit of the perturbed Hamiltonian 
with the same initial condition as $\gamma$, while $\gamma_V'$ is the orbit of 
the perturbed Hamiltonian with the same initial condition as $\gamma'$. 
The endpoints of $\gamma$  and $\gamma_V'$ are separated by less
than a quantum-mechanical
resolution scale (shaded area). Right panel: phase space displacement. 
Labels are the same as in the left panel, with $\bf{P}$ replacing $V$ as 
subscript for perturbed trajectories. 
Note that $\gamma_{\bf{P}}^{\prime}$ and $\gamma$ lie on top of each other.}
\label{structural_figure}
\vspace{-3mm}
\end{figure}

In the case of a uniform phase-space displacement,
the diagonal approximation becomes even better. 
This is so because any classical 
trajectory of the unperturbed 
Hamiltonian is also a trajectory of the perturbed Hamiltonian,
up to displacements at the trajectory's ends.
This is illustrated in the right panel of Fig.~1.
The fact that the action
difference is here time-independent has the important consequence that 
the FGR decay is replaced by 
a time-independent saturation term. The Lyapunov decay term is left almost
unaffected, as it 
depends on the classical measure of nearby trajectories with perturbed 
initial conditions
and does not depend on quantum action phases. We also note that for 
displacement echoes
there is no Gaussian perturbative decay, since phase space displacements 
do not change the spectrum of the system aside from some possible but 
irrelevant global shift.

For a quantitative approach to the problem, we semiclassically evaluate
$M_{\rm D}(t)$ for the case of an initial Gaussian 
wavepacket, $
\bracket{\mathbf{r}}{\alpha \parent{\mathbf{r}_0,\mathbf{p}_0}}=\parent{\pi \sigma^2}^{-\frac{d}{4}}
\exp[i\mathbf{p}_0 \cdot \parent{\mathbf{r}-\mathbf{r}_0}-\parent{\mathbf{r}-\mathbf{r}_0}^2/2 \sigma^2].$
Following Ref.~\cite{heller-propagator}, 
we semiclassically propagate $|\alpha \rangle$ with the help of
the van Vleck propagator, linearly expanding around $\bf{r}_0$:
\begin{eqnarray}
&&\bra{\mathbf{r'}}e^{-i\hat{H}t}\ket{\alpha}_{\rm sc} \simeq 
\parent{-\frac{i\sigma}{\sqrt{\pi}}}^{\frac{d}{2}} \\
&&\times \sum_\gamma \sqrt{C_\gamma}  
\exp[i S_\gamma-i \pi \nu_\gamma/2-\sigma^2 \parent{\mathbf{p}_\gamma-\mathbf{p}_0}/2].\nonumber 
\end{eqnarray}
Here, the sum runs over all possible classical trajectories $\gamma$ connecting
$\mathbf{r}_0$ and $\mathbf{r'}$ in the time $t$, 
$\mathbf{p}_\gamma=\left. -\partial S_\gamma / \partial \mathbf{r}\right|_{\mathbf{r}=\mathbf{r}_0}$
is the initial momentum on $\gamma$, $S_\gamma$ is the classical action accumulated
on $\gamma$, $\nu_\gamma$ is the Maslov index and 
$C_\gamma=\abs{-\partial^2 S_j\parent{\mathbf{r},\mathbf{r}';t} / \partial r_i\partial r^{\prime}_\gamma}_{\mathbf{r}=\mathbf{r}_0}$. The kernel $I(t)$ of $M_{\rm D}(t)$ 
involves a double sum over classical trajectories, $\gamma$ and 
$\gamma'$, and can be interpreted as the overlap between a wavepacket that is boosted and 
subsequently propagated with a wavepacket that is first propagated and subsequently  
boosted~\cite{ins}. Enforcing a stationary phase condition kills all but the contributions 
with the smallest actions. They correspond to $\gamma=\gamma'$ and one has
\begin{eqnarray}\label{kernel}
I\parent{t}& =&\parent{\frac{\sigma^2}{\pi}}^{\frac{d}{2}}
\int {\rm d} \mathbf{r}' \sum_\gamma e^{i\mathbf{P} \cdot (\mathbf{r}'+\mathbf{r}_0)} \, C_\gamma 
\\
& & \times \exp-\frac{\sigma^2}{2}\Big[ \parent{\mathbf{p}_\gamma-\mathbf{p}_0}^2+\parent{\mathbf{p}_\gamma-\mathbf{p}_0-\mathbf{P}}^2 \Big].\nonumber 
\end{eqnarray}
Taking the squared amplitude $|I(t)|^2$ one obtains the semiclassical expression for the
displacement echo
\begin{eqnarray}
M_{\rm D} (t ) &=& 
\left(\frac{\sigma^2} {\pi}\right)^{d} \int {\rm d}{\bf r} \, {\rm d}{\bf r}^{\prime} 
 \, \sum_{\gamma,\gamma'}e^{{\it i} {\bf P} \cdot ({\bf r} -{\bf r}^{\prime})} \,
 C_{\gamma}\,
 C_{\gamma^{\prime}}\, \\
& \times &\exp -\frac{\sigma^2}{2}\Big[\parent{\mathbf{p}_\gamma-\mathbf{p}_0}^2+\parent{\mathbf{p}_\gamma-\mathbf{p}_0-\mathbf{P}}^2 \Big]\nonumber \\
&\times & \exp -\frac{\sigma^2}{2}\Big[
     \parent{\mathbf{p}_{\gamma'}-\mathbf{p}_0}^2+\parent{\mathbf{p}_{\gamma'}-\mathbf{p}_0-\mathbf{P}}^2 
\Big] . \nonumber 
\end{eqnarray}
We calculate $\langle M_{\rm D}(t) \rangle$, 
the ensemble-averaged displacement echo 
over a set of initial 
Gaussian wavepackets with varying center of mass ${\bf r}_0$. 
There are two qualitatively different contributions to 
$\langle M_{\rm D}(t) \rangle$. The first 
contribution $\langle M_{\rm D}(t) \rangle_{\rm c}$
comes from pairs $\gamma \approx \gamma'$ of correlated 
trajectories that remain within a distance $\lesssim \sigma$ of each other 
for the whole duration $t$ of the experiment, while the second contribution
$\langle M_{\rm D}(t) \rangle_{\rm u}$ arises from pairs of uncorrelated 
trajectories $(\gamma,\gamma')$. For the first contribution, we write
$\exp[i {\bf P} ({\bf r}-{\bf r}')] \approx 1$, which is true in the 
semiclassical limit where $\sigma \rightarrow 0$, and set $\gamma=\gamma'$.
One then has
\begin{eqnarray}\label{corr1}
\langle  M_{\rm D}(t)\rangle_{\rm c} &=& 
\left(\frac{\sigma^2} {\pi}\right)^{d} 
\int {\rm d}{\bf r} {\rm d}{\bf r}^{\prime} \, \delta_{\sigma}
({\bf r} -{\bf r}^{\prime}) \nonumber \\
&
\times& \Big\langle \sum_{\gamma} \,
 C_{\gamma}^2\, e^{-\sigma^2\left[ \parent{\mathbf{p}_\gamma-\mathbf{p}_0}^2+\parent{\mathbf{p}_\gamma-\mathbf{p}_0-\mathbf{P}}^2 \right]} \Big\rangle,
\end{eqnarray}
where $\delta_{\sigma} ({\bf r} -{\bf r}^{\prime})$ restricts the integrals to
$|{\bf r} -{\bf r}^{\prime}| \le \sigma$. The calculation of (\ref{corr1})
is straightforward. The integral over ${\bf r}'$ gives a factor $\sigma^d$.
One then replaces one $C_\gamma$ by its asymptotic
value, $C_\gamma \propto \exp[-\lambda t]$, and uses the second $C_\gamma$ to
perform a change of integration variable
$\int {\rm d}{\bf r} \sum_\gamma C_\gamma = \int {\rm d} {\bf p}$. 
After a Gaussian integration, one finally gets the correlated
contribution to $\langle  M_{\rm D}(t)\rangle$ as
\begin{eqnarray}\label{corr2}
\langle  M_{\rm D}(t)\rangle_{\rm c} &=& 
\alpha \, \exp[-({\bf P} \sigma)^2/2] \, \exp[-\lambda t].
\end{eqnarray}
Here, $\alpha$ is a weakly time-dependent number of order one~\cite{jalabert}.

For the uncorrelated part, an ergodicity assumption is justified at 
sufficiently large times, under which one gets
\begin{subequations}
\begin{eqnarray}
\langle M_{\rm D} (t ) \rangle_{\rm u}  
&=& f({\bf P}) \; \langle \tilde{M}_{\rm D}(t) \rangle_{\rm u} , \\
f({\bf P}) &=& \frac{1}{\Omega^2} \int {\rm d}{\bf r} {\rm d}{\bf r}^{\prime} 
\exp[i {\bf P} \cdot ({\bf r} -{\bf r}^{\prime})] , \\
\langle  \tilde{M}_{\rm D}(t)\rangle_{\rm u} &=& \left( \frac{\sigma^2} {\pi}\right)^{d} 
\Big( \int {\rm d}{\bf x} 
 \, \sum_{\gamma} \,
 C_{\gamma}\, \\
& \times & \exp -\frac{\sigma^2}{2}\Big[ \parent{\mathbf{p}_\gamma-\mathbf{p}_0}^2+\parent{\mathbf{p}_\gamma-\mathbf{p}_0-\mathbf{P}}^2 \Big] \Big)^2,\nonumber 
\end{eqnarray}
\end{subequations}
with the system's volume $\Omega \propto L^d$. 
It is straightforwardly seen that 
$\langle  \tilde{M}_{\rm D}(t)\rangle_{\rm u} = \exp[-({\bf P} \sigma)^2/2]$,
and $f({\bf P})=g(|{\bf P}|L)/(|{\bf P}|L)^2$, in term of
an oscillatory function $g(|{\bf P}|L)=4 \sin^2(|{\bf P}|L/2)$ for $d=1$
and $g(|{\bf P}|L)=4 J_1^2(|{\bf P}|L)$ for $d=2$. For $d=3$, 
$g$ is given by Bessel and Struve functions. The uncorrelated contribution
to the displacement echo reads
\begin{eqnarray}
\langle  M_{\rm D}(t)\rangle_{\rm u} &=& 
\exp[-({\bf P} \sigma)^2/2] \; g(|{\bf P}|L)\Big/(|{\bf P}| L)^2,
\end{eqnarray}
which, together with Eq.~(\ref{corr2}) gives the average displacement echo 
as
\begin{eqnarray}\label{corrtot}
\langle  M_{\rm D}(t)\rangle =
\exp[-({\bf P} \sigma)^2/2] \, \left [\alpha \, e^{-\lambda t} \, +
\frac{g(|{\bf P}|L)}{(|{\bf P}| L)^2} \right].
\end{eqnarray}
In addition,
as is the case for Loschmidt echoes, $\langle  M_{\rm D}(t)\rangle
\ge N^{-1}$ where $N$ is the size of Hilbert space.

Eq.~(\ref{corrtot}) is our main result. It states that $M_{\rm D}(t)$ is
the sum of a time-dependent decaying term of classical origin and a
time-independent term of quantum origin. The latter can also be obtained 
within random matrix theory. The prefactor
$\exp[-({\bf P} \sigma)^2/2] \rightarrow 1$ in the semiclassical limit and
is thus of little importance. We see that generically, 
$M_{\rm D}(t)$ follows a classical exponential decay, 
possibly interrupted by a quantum freeze as long as the displacement
is not too large, $g(|{\bf P}|L)\Big / (|{\bf P}| L)^2 > N^{-1}$ \cite{prosen}.
We note that in the semiclassical limit,
$M_{\rm D}(t\rightarrow 0) \rightarrow 1$, 
because of the saturation of 
$\alpha(t \rightarrow 0) \rightarrow 1$
and the disappearance of 
uncorrelated contributions at short times.
Most importantly, there is no displacement-dependent decay, i.e. no counterpart to the FGR decay nor the perturbative Gaussian decay for $M_{\rm D}(t)$, because phase-space displacements 
leave the spectrum unchanged, up to a possible homogeneous shift~\cite{doron-alex}.

\begin{figure}
\begin{center}
\includegraphics[width=6.5cm]{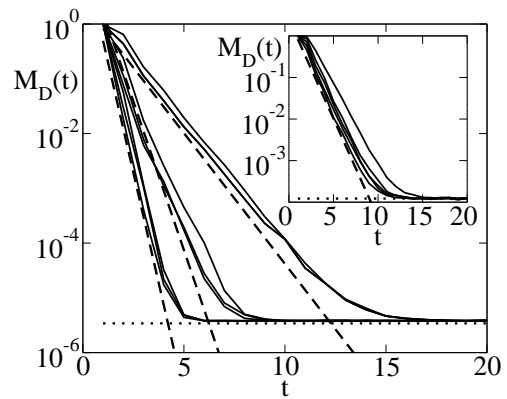}
\end{center}
\vspace{-3mm}
\caption{Main plot : Displacement echo $M_{\rm D}(t)$ for the kicked rotator 
model with $N=262144$,
and displacements $P= m \times 2 \pi/N$, $m=10,\, 20, \,30$.
Averages have been performed over a set of 10000 different initial coherent
states. The full lines correspond to kicking strengths $K= 10.09 $, 50.09 
and $200.09$ (from right to left)
and the dashed lines (slightly shifted for clarity) 
give the predicted exponential decay given by the reduced
Lyapunov exponent  $\lambda_0 = 1.1,\, 2.5,\, 3.7$.
The dotted line gives the saturation at $N^{-1}.\,$
Inset : Displacement echo for $N=8192$, $K= 10.09 $, and 
displacements $P=2 \pi/N,\, 4 \pi/N, \ldots 10 \pi/N$.
Data are obtained from 1000 different initial coherent states.
The dashed line gives the predicted exponential decay given by the reduced 
lyapunov exponent  $\lambda_0 = 1.1$.
The dotted line gives the minimal saturation value at $N^{-1}.\,$}
\vspace{-3mm}
\end{figure}

What does the ``freeze'' correspond  to physically? 
It is the elastic component in any of the mentioned spectroscopies: 
M\"ossbauer, neutron, and molecular electronic, and was first identified by 
van Hove in connection with neutron scattering~\cite{vanhove}.  
There is a finite probability, above the $1/N$ statistical limit, 
of not changing quantum state in spite of being ``hit''; 
this is the source for example of 
the recoilless peak in M\"ossbauer spectroscopy.

We now check our  predictions numerically . 
We specialize to the kicked rotator model with Hamiltonian
\begin{equation}\label{kickrot}
H_0 = \frac{\hat{p}^2}{2} + K \cos \hat{x} \sum_n \delta(t-n).
\end{equation}
We focus on the regime $K > 7$, for which the dynamics is fully
chaotic with Lyapunov exponent $\lambda = \ln[K/2]$. 
We quantize this Hamiltonian on a torus, which 
requires to consider discrete values
$p_l=2 \pi l/N$ and $x_l=2 \pi l/N$, $l=1,...N$. In these units,
one has $L=N$. The displacement
echo of Eq.~(\ref{decho}) is computed for discrete times $t=n$, as
\begin{eqnarray}\label{krot_fid}
M_{\rm D}(n) = \big|\langle \psi_0 | e^{-i P \hat{x}}
\left({\cal U}^\dagger\right)^n 
e^{i P \hat{x}} \left({\cal U}\right)^n  | \psi_0 \rangle \big|^2,
\end{eqnarray}
with $P=|{\bf P}|$. Here, we used the unitary Floquet time-evolution
operator ${\cal U}$ whose 
matrix elements in $x-$representation are given by
\begin{eqnarray}
{\cal U}_{l,l'} & = & \frac{1}{\sqrt{N}} \exp\Bigg[i 
\frac{\pi (l-l')^2}{N}\Bigg] \exp\Bigg[-i \frac{N K}{2 \pi} \cos \frac{2 \pi l'}{N}\Bigg].
\nonumber
\end{eqnarray}
The time-evolution of $\psi_0$ in
Eq.~(\ref{krot_fid}),
is calculated by recursive calls to a fast-Fourier transform routine.

Fig.~2 shows the behavior of the echo for displacements in the
range $P \gg 2 \pi/N$ 
for which $\langle  M_{\rm D}(t)\rangle_{\rm u} \ll N^{-1}$
and thus plays no role. It is seen that the decay rate 
of the displacement echo strongly depends on the kicking strength $K$, but
is largely independent of the displacement $P$. We quantitatively found
that in that regime, $M_{\rm D}(t) \approx \exp[-\lambda_0 t]$, in term of the
reduced Lyapunov exponent $\lambda_0$ \cite{caveat2}. The inset shows
moreover, that lowering the displacement to the regime $P=m 2\pi/N$ 
with $m \le 5$ does not affect the decay rate of $M_{\rm D}(t)$, i.e.
there is no FGR decay for the displacement echo.

We focus in Fig.~3 on smaller displacements
$P \le  2 \pi / N$. The behavior of $\langle M_{\rm D}(t) \rangle$ clearly
satisfies 
Eq.~(\ref{corrtot}), with a quantum freeze at a displacement-dependent
value following a decay with a slope given by the Lyapunov exponent. We show in
the main panel the $P$-dependence 
of the value at which $M_{\rm D}(t)$ freezes. The data unambiguously confirm 
the algebraically damped oscillations
predicted in Eq.~(\ref{corrtot}).

\begin{figure}
\begin{center}
\includegraphics[width=6.5cm]{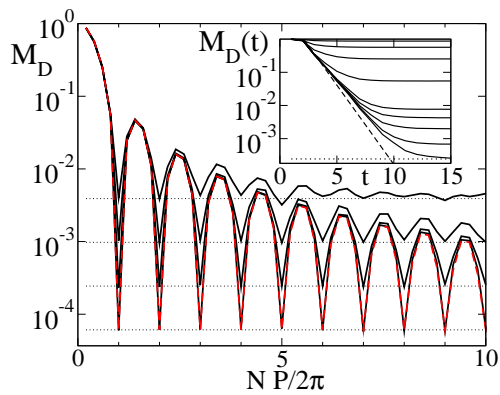}
\end{center}
\vspace{-3mm}
\caption{(color online) 
Main plot: Saturation value $M_{\rm D}(\infty)$ of the displacement echo
as a function of the rescaled displacement 
$N P/2 \pi$ for the kicked rotator 
model with $N = 256$,  $ 1024$,  $ 4096$, $16384$ (full lines, from 
top to bottom). Data are obtained from 1000 different initial coherent states.
The dotted lines give the saturation at $N^{-1}.\,$ The red dashed line 
gives the theoretical prediction 
$M_{\rm D}(\infty) = 
{\rm Max} (4 \exp[-(\sigma P)^2/2] \sin^2(P L/2)\Big/(PL)^2 , N^{-1})$ for $N=16384$.
Inset: Quantum freeze of the displacement echo for kicking strength 
$K=10.09$, $N=4096$, and $P\in[0,2 \pi/N]$.
The dashed line gives the decay with the reduced Lyapunov exponent
$\lambda_0 = 1.1$ (see text).}
\vspace{-3mm}
\end{figure}

In summary, we have presented a semiclassical calculation of 
phase-space displacement echoes. We showed that 
they are generically given by the sum of 
a classical decay and a quantum freeze term (\ref{corrtot}). 
Because phase-space displacements do not generate time-dependent
action differences, and because they vanish in
first order perturbation theory, there is no other time-dependent
decay, in contrast to Loschmidt 
echoes \cite{peres,jalabert,jacquod,tomsovic,vanicek}. 

To conclude, we note that 
neutron scattering correlation functions are given by the 
average $\langle I(t) \rangle$ of the kernel of $M_{\rm D}(t)$.
Starting back from Eq.~(\ref{kernel}), one gets
 \begin{eqnarray}
|Y_{jj}\parent{\bf{P},t}| \simeq 
\exp[-({\bf P} \sigma)^2/4] \,
\frac{g^{1/2}(|{\bf P}|L)}{|{\bf P}| L},
\end{eqnarray}
i.e. $|Y_{jj}|$ is given by the quantum freeze term only. 
This is so, since the correlations between pairs of trajectories
that are necessary for the existence of the Lyapunov term appear only
once $I(t)$ is squared. 

C. Petitjean was supported by the Swiss NSF, and
D. V. Bevilaqua was partially supported by CAPES, Coordena\c{c}\~{a}o de 
Aperfei\c{c}oamento de 
Pessoal de N\'{i}vel Superior and by NSF grant no. CHE-0073544.


\begin{thebibliography}{99}
\bibitem{caveat1} More generally, any
measurement of momentum or position time correlators (or combinations of 
the two) can be viewed as a fidelity experiment under certain 
phase space displacement.
\end{thebibliography}
\end{document}